\newif\ifOneCol

\ifOneCol
\documentclass[draftcls,12pt,onecolumn]{IEEEtran}
\else
\documentclass[journal]{IEEEtran}
\fi
\IEEEoverridecommandlockouts
\usepackage{mathtools}
\usepackage{graphicx}
\usepackage[noadjust]{cite}
\usepackage{amsmath}
\usepackage{textcomp}
\usepackage{tikz}
\usepackage[caption=false]{subfig}
\usepackage{array}
\usepackage{bm}
\usepackage{algorithm, algorithmicx}
\usepackage{algpseudocode}
\usepackage{mathrsfs}
\usepackage{amssymb}
\usepackage{enumerate}
\usepackage{ragged2e}
\usepackage{filecontents}
\usepackage{url}
\usepackage{times}
\usepackage{multirow}
\usepackage{stfloats}
\usepackage[shortlabels]{enumitem}

\def\harq{{\text{HARQ}}}

\def\switch{{\text{switch}}}

\def\rep{{\text{rep}}}

\def\tb{{\text{TB}}}
\def\data{{\text{data}}}

\def\hc{{\text{HC}}}
\def\suf{{\text{SUF}}}

\def\rtt{{\text{RTT}}}
\def\ack{{\text{ACK}}}
\def\pusch{{\text{PUSCH}}}
\def\pdsch{{\text{PDSCH}}}
\def\pucch{{\text{PUCCH}}}
\def\pdcch{{\text{PDCCH}}}

\ifCLASSINFOpdf

\else

\fi

\allowdisplaybreaks
\begin{document}
	
	\title{Enhanced Hybrid Automatic Repeat Request Scheduling for Non-Terrestrial IoT Networks}
	
	\author{Gautham Prasad, Vishnu Rajendran Chandrika, Lutz Lampe, \textit{Senior Member, IEEE}, Gus Vos, \textit{Senior Member, IEEE} 
		\thanks{A part of this paper was presented at the 3GPP TSG-RAN meeting \#104e, Jan.-Feb., 2021\textit{}~\cite{R1_2101323}.}
		\thanks{Gautham Prasad is with Ofinno, Reston, VA. Vishnu Rajendran and Lutz Lampe are with the Department of Electrical and Computer Engineering, The University of British Columbia, Vancouver, BC, Canada. Gus Vos is with Sierra Wireless Inc., Richmond, BC, Canada. Email: gautham.prasad@alumni.ubc.ca, vishnurc@ece.ubc.ca, lampe@ece.ubc.ca, gvos@sierrawireless.com. }
		\thanks{This work was supported by the Natural Sciences and Engineering Research Council of Canada (NSERC) and Sierra Wireless.}
	}
	
\markboth{submission to an IEEE journal}%
{Submitted paper}

\maketitle
	
\begin{abstract}
Non-terrestrial networks (NTNs) complement their terrestrial counterparts in enabling ubiquitous connectivity globally by serving unserved and/or underserved areas of the world. \textcolor{black}{Supporting enhanced mobile broadband (eMBB) data over NTNs has been extensively studied in the past. However, focus on massive machine type communication (mMTC) over NTNs is currently growing. Evidence for this are the work items included into the 3rd generation partnership project (3GPP) agenda for commissioning specifications for Internet-of-Things (IoT) communications over NTNs.} Supporting mMTC in non-terrestrial cellular IoT (C-IoT) networks requires jointly addressing the unique challenges introduced in NTNs and C-IoT communications. In this paper, we tackle one such issue caused due to the extended round-trip time and increased path loss in NTNs resulting in a degraded network throughput. We propose smarter transport blocks scheduling methods that can increase the efficiency of resource utilization. \textcolor{black}{We conduct end-to-end link-level simulations of C-IoT traffic over NTNs. Our numerical results of data rate show the improvement in performance achieved using our proposed solutions against legacy scheduling methods.}

\end{abstract}
	
	
	\IEEEpeerreviewmaketitle
	
\section{Background}\label{sec:intro}
\IEEEPARstart{N}on-terrestrial networks (NTNs), including those enabled by satellites in the low earth orbit (LEO), medium earth orbit (MEO), and geostationary earth orbit (GEO), as well as high-altitude platform stations (HAPS) and other unmanned/uncrewed aerial vehicles (UAVs), complement their conventional terrestrial counterparts in enhancing cellular coverage by serving unserved and/or underserved areas~\cite{9149179, 9107401}. \textcolor{black}{NTNs are critical in remote areas with low/no cellular connectivity. They are applicable across many different industries, such as transportation (maritime, road, rail, and air) and logistics, farming, mining, utilities, and environment monitoring~\cite{lee2020integrating, 8911669, lin20195g}.} As a result, fifth generation (5G) and beyond (B5G) cellular networks are increasingly focusing on supporting NTNs for both enhanced mobile broadband (eMBB) and low power wide area network (LPWAN) applications~\cite{tr_38821, R1_2008868, mahmood2020white}. 

Adapting cellular communications in NTNs requires addressing the several challenges introduced in the new system setting. The separation distance between the user equipment (UE) and the base station (BS) is increased in NTNs where the BSs are, or are connected to the UEs via, satellites or UAVs\footnote{Throughout the rest of the paper, we refer to an NTN payload on an airborne platform as satellite as it is the most common type of platform. However, our descriptions, analyses, and evaluations are also valid when the platform is a UAV, or a drone, or any other non-terrestrial platform entity.}. This results in large propagation delay and increased path loss of the transmitted signal. Furthermore, NTNs are associated with BSs or relay terminals that move at high velocities, which also impose challenges such as varying propagation time and increased impact of Doppler effects. \textcolor{black}{Solutions to counter these effects have been proposed in the past for eMBB applications~\cite{tr_38821}. However, methods for incorporating cellular Internet-of-Things (C-IoT) networks into the fold of NTN are still in their infancy~\cite{tr_36763}.}

Support for C-IoT and massive machine type communication (mMTC) systems in NTNs is critical across several domains. \textcolor{black}{NTN serves C-IoT scenarios in both wide- and local-area IoT services. The former includes gathering data on a macro-level from sensors deployed across broad geographical areas. Automotive support (e.g., enabling over-the-air auto upgrades, vehicle platooning, traffic flow optimization), large-scale infrastructure monitoring in energy distribution systems, and managing livestock and farming in the agriculture industry form the core of wide-area IoT application scenarios~\cite{tr_38811}. On the other hand, sensors and actuators serving a more confined area, such as on board a maritime vessel or a neighborhood advanced metering infrastructure in a smart grid sub-system, make up local-area IoT networks~\cite{tr_38811}. As with eMBB applications, NTN allows for expanding IoT networks, both wide- and local-area services, in a ubiquitous manner and guaranteeing continuity of service across geographical areas~\cite{liberg2020narrowband}.}

In this paper, we propose methods for enhancing resource scheduling tailored for such C-IoT and mMTC systems to operate in NTNs, which are collectively referred to as \textit{IoT-NTN}. In particular, we focus on the impact of the extended round trip time (RTT) of bidirectional signals caused due to the increased propagation distance in NTNs on the network throughput. \textcolor{black}{Typically, the achievable data rate may not be the primary target performance indicator to focus on while designing C-IoT and mMTC systems. One of the reasons for this is the greater resource consumption (e.g., extended bandwidth requirement) and higher transmit power associated with achieving increased throughput. However, we show in this work that the network throughput can be improved without demanding additional resources or increasing transmit power, but instead by efficiently scheduling the hybrid automatic repeat request (HARQ) processes.} As a result, the IoT-NTN can support a larger number of IoT devices, which is critical in NTNs due to the significantly larger cell size compared to conventional terrestrial networks. \textcolor{black}{Toward this end, we exploit the lengthy RTT in NTN links to overlap bidirectional signals in the air. We further utilize the frequency-division full-duplex nature of BSs to allow simultaneous transmission and reception at the satellites and BSs.}

The major contributions of our paper can be listed as follows:
\begin{itemize}
    \item We propose HARQ scheduling designs for IoT-NTN systems to increase the uplink (UL) and downlink (DL) resource utilization efficiency and throughput. To achieve this, we develop flexible DL data to acknowledgement (DD2A) delays and UL grant to data (UG2D) delays for more efficient bidirectional signaling. We ensure that our solution works for different RTTs observed in NTNs, and also with the usage of a varied number of HARQ processes.
    \item We propose methods to signal and calculate DD2A and UG2D delays for dynamic transmission adaptation conditions, e.g., changing adaptive modulation and coding schemes (MCS) and varying RTTs. We address both scenarios where the number of transport block repetitions are the same and different within a given scheduling interval to render to our solution future-proof.
    \item We provide a thorough investigation and simulation-based evaluation for determining the transport block size (TBS) and forward error correction strategy for improving spectral efficiency under varying IoT-NTN link conditions. Using these settings, we also present the evaluation results of our proposed solutions in terms of the achieved data rate gains. 
\end{itemize}

\textit{Outline}: The rest of the paper is organized as follows. \textcolor{black}{In Section~\ref{sec:system_model}, we first introduce the system model, and use the metrics and key performance indicators from the model to highlight the shortcomings of relevant prior works in tackling the issue we address in this paper.} We propose our solutions in Section~\ref{sec:proposed_solutions}, which we evaluate and present their performance results in Section~\ref{sec:results}. We conclude the paper in Section~\ref{sec:conclusion}. We also include a list of important notations and their meanings in Table~\ref{table:notations}.
\begin{table*}[t]
	\centering
	\caption{List of important notations used in the paper}\label{table:notations}
	\begin{tabular}{ll|ll}
		\hline
Notation           & Meaning                                                               & Notation                & Meaning 
\\ \hline \hline
$\eta_\suf$        & SF utilization   factor                   & $n_{\text{DD2A}}$       & Delay between the DL data and the   associated ACK in units of SFs    \\
$N_\data$          & Number of DL or UL data SFs               & $n_{\text{UG2D}}$       & Delay between the UL grant and the associated UL data in units of SFs \\
$N_\rep$           & Number of repetitions of TB               & $N_{\text{DG2D}}$       & Delay between the DL grant and the associated DL data in units of SFs \\
$N_\hc$            & Lengths of HARQ cycle in units of SFs     & $n_{\text{DD2A, min}}$  & Minimum mandatory DD2A in units of SFs                                \\
$R$                & Data rate                                 & $n_{\text{UG2D, min}}$  & Minimum mandatory UG2D in units of SFs                                \\
$n_\tb$            & Transport block size in bits              & $n_{\text{bundle}}$     & Number of ACKs bundled within one TTI                                 \\
$t_\tb$            & Transport block duration in seconds       & $\gamma$                & Signal-to-noise ratio                                                 \\
$N_\harq$          & Number of HARQ processes                  & $\beta_{\text{PL}}$     & Free space path loss                                                  \\
$\tau_\rtt$        & RTT  in  seconds                          & $\delta$                & Signal bandwidth                                                      \\
$n_{\rep,\data}$    & Number of data block repetitions          & $P_{\text{EIRP}}$       & Effective isotropically radiated power of UE                          \\
$n_{\rep, \pdsch}$ & Number of PDSCH repetitions               & $G/T$                   & Satellite antenna-gain-to-noise-temperature                           \\
$n_{\rep, \pdcch}$ & Number of PDCCH repetitions               & $k$                     & Boltzmann constant                                                    \\
$n_{\rep, \pucch}$ & Number of PUCCH repetitions               & $f$                     & Carrier frequency                                                     \\
$n_{\rep, \pusch}$ & Number of PUSCH repetitions               & $\beta_{\text{atm}}$    & Atmospheric loss                                                      \\
$N_\switch$        & Switching delay in units of SFs           & $\beta_{\text{shadow}}$ & Shadow fading margin                                                  \\
$N_{\text{TBPHC}}$ & Number of TBs scheduled in one HARQ cycle & $\beta_{\text{scint}}$  & Scintillation loss                                                    \\
$N_{\text{A2G}}$   & ACK processing delay in units of SFs      & $\beta_{\text{polar}}$  & Polarization loss                                                 \\ \hline          
\end{tabular}
\end{table*}

\section{\textcolor{black}{Prerequisites}} \label{sec:system_model}
\subsection{\textcolor{black}{System Model}}
\begin{figure}[t]
	\centering
	\includegraphics[width=8cm]{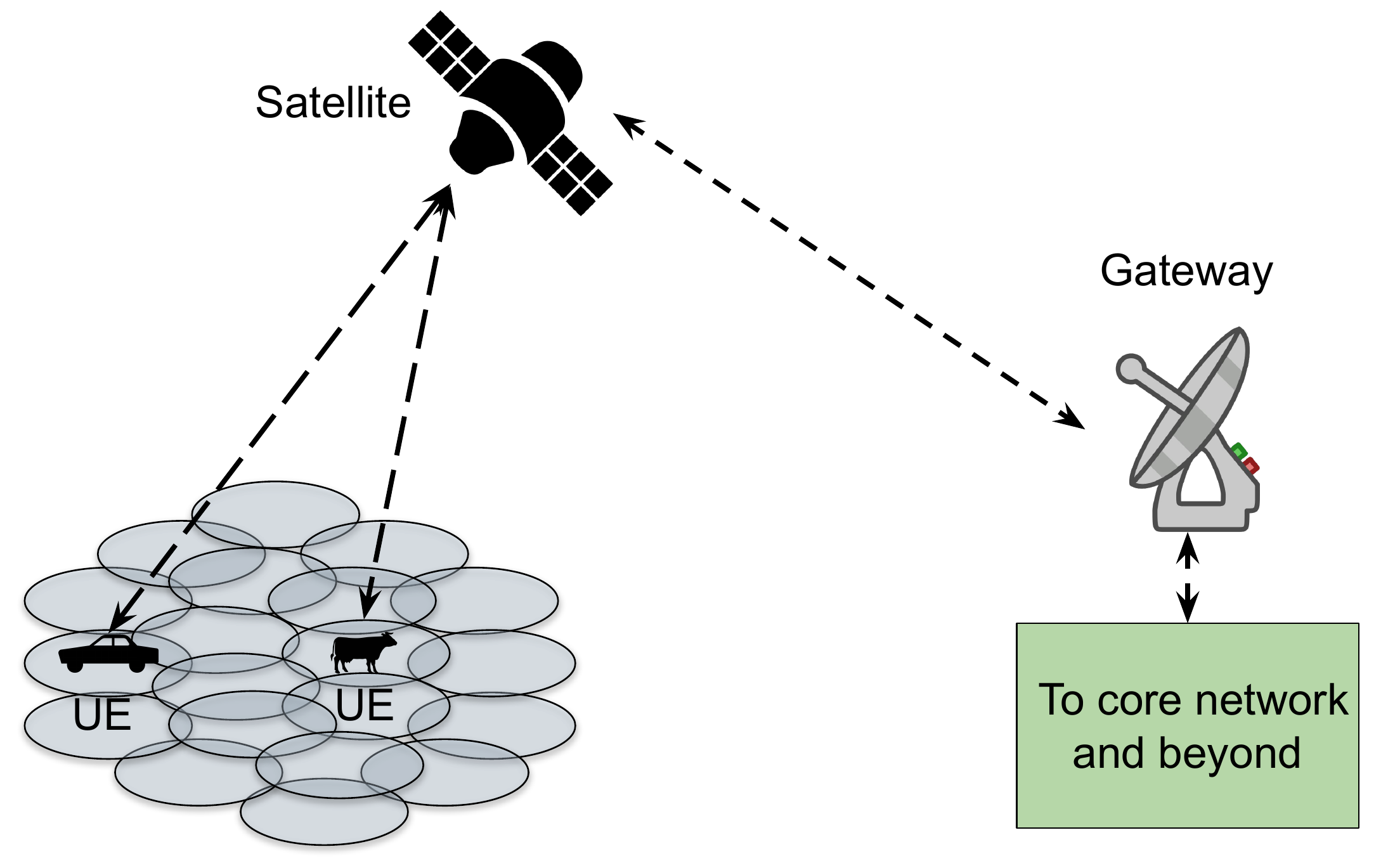} 
	\caption{\textcolor{black}{An illustration of an IoT-NTN system with terrestrial UEs being served by a non-terrestrial satellite.}}
	\label{fig:NTNsystem}
\end{figure}
\textcolor{black}{We show the overall system model of a typical IoT-NTN in Fig.~\ref{fig:NTNsystem}. In this work, we consider the satellite to be in a LEO revolving around the earth in a circular orbit. But our study is directly applicable to higher satellite altitudes of MEO and GEO satellites and also for lower altitudes of HAPS and UAVs. Our study is agnostic to the data processing/forwarding architecture, and thus supports both transparent or bent-pipe payloads and regenerative ones~\cite{tr_38821, rinaldi2020non}. Our proposed solutions are also applicable to both earth-moving and earth-fixed NTN cell-types~\cite{geraci2022integrating}.}

\begin{table}[t]
	\centering
	\caption{RTTs for LEO NTNs} \label{table:rtt}
	\begin{tabular}{c|c|c|c}
		\hline 
		Satellite Altitude & Payload Type & Min. RTT (ms) & Max. RTT (ms) \\ 
		\hline \hline
		600 km & Regenerative & 4 & 13 \\
		\hline
		600 km & Transparent & 8 & 26 \\
		\hline
		1200 km & Regenerative & 8 & 21 \\
		\hline
		1200 km & Transparent & 16 & 42 \\
		\hline
	\end{tabular}
\end{table}

\textcolor{black}{An illustration of our considered system architecture is shown in Fig.~\ref{fig:NTNsystem}. The beam footprint of each beam determines the RTT range of the satellite link~\cite{tr_38821}.} We present the RTTs of IoT-NTN links for different LEO altitudes for both types of payload architectures in Table~\ref{table:rtt}. The minimum and the maximum RTTs correspond to the maximum ($90^\circ$) and minimum ($10^\circ$) beam elevation angles, respectively.

\begin{figure}[t]  
	\centering
	\subfloat[]
	{\includegraphics[width=8cm]{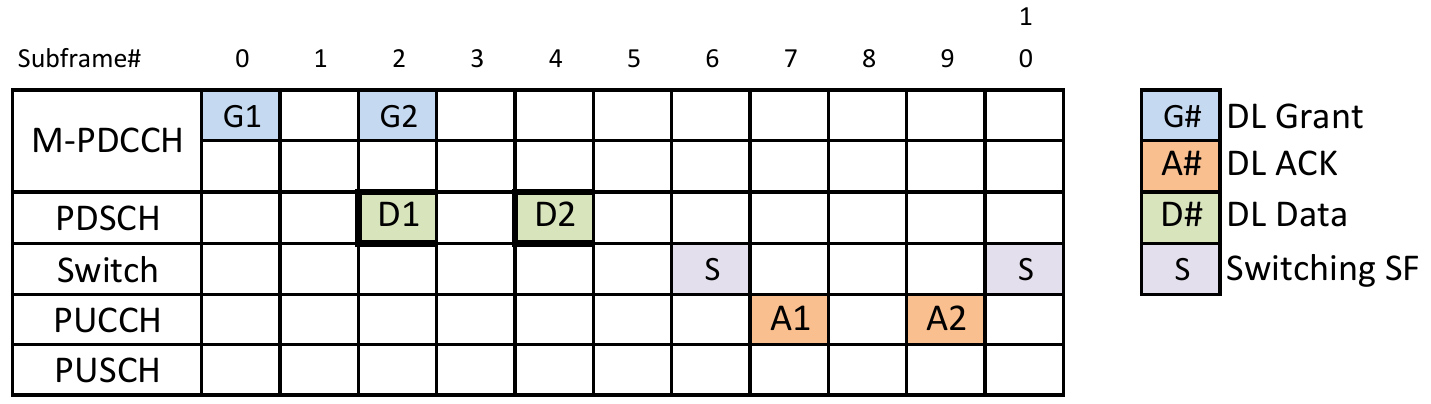}}   \\
	\subfloat[]
	{\includegraphics[width=7cm]{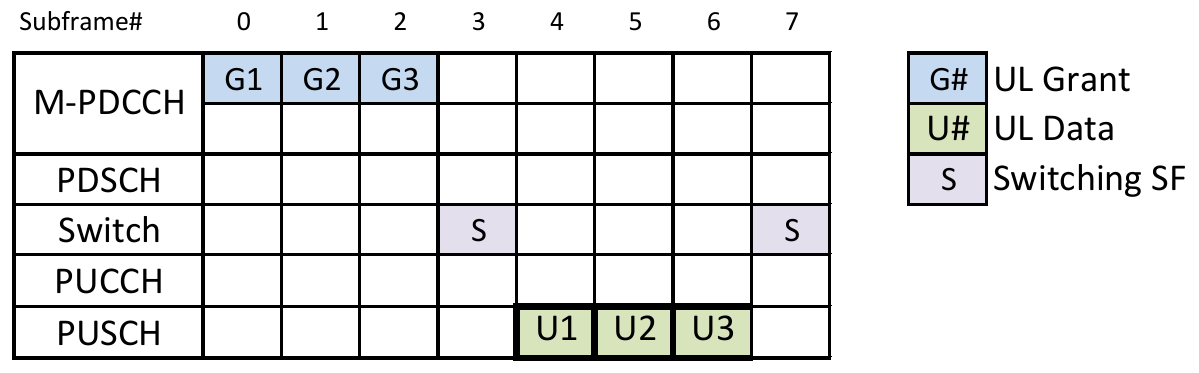}}
	\caption{{Timing diagrams for LTE-M operation in (a) the DL and (b) the UL.} }  \label{fig:legacy_timing}
\end{figure}

We borrow the UL and DL communication mechanism between the UE and BS for IoT-NTN directly from the long term evolution - MTC (LTE-M) and narrowband IoT (NB-IoT) specifications~\cite{ts_36321, ts_36213}. \textcolor{black}{Our reasoning behind this is two-fold. First, LTE-M and NB-IoT are the industry chosen standards for enabling low-power wide area networks. Several works have been presented in the past to demonstrate the benefits of using these 3GPP standards, e.g.,~\cite{sorensen2022modelling, aldahdouh2019survey}. The use of LTE-M and NB-IoT provides the advantage of reusing existing infrastructure and operating on licensed bands. These help in providing low-cost, stable, reliable, and predictable performance across application scenarios. Second, similar to using the new radio (NR) standard as the starting point for eMBB over NTN~\cite[Sec. 6]{tr_38821}, IoT-NTN standardization activities in 3GPP have agreed to build on the existing LTE-M and NB-IoT standards to expand them into the NTN realm~\cite{tr_36763}. By showing the effectiveness of our proposed solution for LTE-M and NB-IoT, we aim to demonstrate that integrating our cost-efficient method into legacy systems is practically feasible. According to both the LTE-M and NB-IoT standards, UL and DL data bits are grouped into transport blocks (TBs) of varying sizes for transmission~\cite{ts_36213}.} The TBS is dependent on the adaptive MCS chosen based on the operating conditions and target block error rate (BLER). The transmission time interval (TTI), which is the time spanned by one unit of transmission corresponds to one sub-frame (SF) of $1$~ms duration, during which one or more TBs are transmitted. A timing diagram of the DL and UL transmissions for LTE-M are shown in Figs.~\ref{fig:legacy_timing}(a) and (b), respectively. Figs.~\ref{fig:legacy_timing}(a) and (b) demonstrate the HARQ-based DL and UL communication~\cite[Ch. 10]{sesia2011lte}, which is used in both LTE-M and NB-IoT. \textcolor{black}{The need for using a HARQ-based design in cellular low-power wide area network technologies, such as LTE-M and NB-IoT, has been extensively shown in the past, e.g.,~\cite{wang2017primer}.}

A DL data TB on the physical downlink shared channel (PDSCH) is preceded by a corresponding single TB grant (STBG) sent on the MTC physical data control channel (M-PDCCH). For the case of NB-IoT, the grants are sent on the NB-IoT PDCCH (N-PDCCH). Henceforth, we drop the prefix for brevity and refer to it only as PDCCH. One grant may also configure multiple TBs, and such grants are referred to as multiple TB grants (MTBGs). \textcolor{black}{For every data TB received, the UE responds with an acknowledgment (ACK) TB on the physical uplink control channel (PUCCH). The ACK TB can either acknowledge one or more TBs using unbundled or bundled ACKs, respectively.} UL transmissions are analogous to DL, where UL data TBs that are configured by UL grants (UGs) in an STBG or MTBG fashion are transmitted on the physical uplink shared channel (PUSCH) by the UE. For this work, we consider low-cost IoT UEs that operate in half-duplex and frequency division duplexed (HD-FDD) manner, which is the industry preferred design for cost and complexity reduction~\cite{hoglund2018overview, BORKAR2020145, rajendran2020scuba}. As a result, the UE uses one or more SFs to switch between transmission and reception modes. 

\begin{figure*}[t]
	\centering
	\includegraphics[width=15cm]{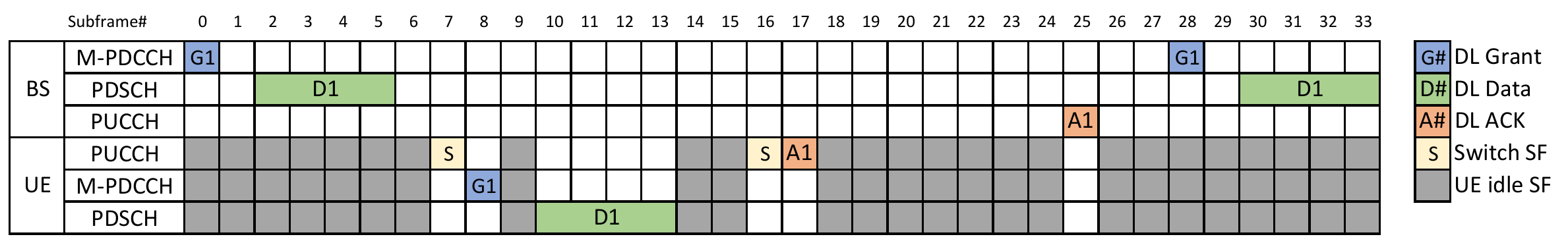} 
	\caption{Timing diagram for DL operation in NTN.}
	\label{fig:ntn_mtc_harq}
\end{figure*}

The timing diagrams shown in Fig.~\ref{fig:legacy_timing} are identical at both the BS and UE for terrestrial networks due to the negligible signal propagation delay. \textcolor{black}{However, in NTNs, the timing shown in Fig.~\ref{fig:legacy_timing} is valid at the UE, while the DL transmissions are sent in advance from the BS by half the RTT. Similarly, the UL packets reach the BS also after half the RTT.} We show an example of this in the timing diagram of Fig.~\ref{fig:ntn_mtc_harq}. We see that the overhead caused by the processing delays and the time of flight between the UE and BS results in several idle SFs at the UE-end, which significantly reduce the achievable throughput.

To quantify the data rates obtainable, we define the concepts of HARQ cycle and SF utilization factor (SUF). We define one HARQ cycle as the total duration of time during which the UE receives on the PDCCH and/or PDSCH and the duration of time it transmits on PUCCH or PUSCH. A HARQ cycle also includes the switching SF(s). For example, Fig.~\ref{fig:legacy_timing}(a) and Fig.~\ref{fig:legacy_timing}(a) show one HARQ cycle for DL and UL, where the lengths of a HARQ cycle, $N_\hc$, in units of SFs are $N_\hc = 11$ and $N_\hc=8$, respectively. 

Next, we define the SUF, $\eta_\suf$, as
\begin{equation}\label{eq:suf}
  \eta_\suf = \frac{N_\data}{n_\rep  N_\hc},
\end{equation}
where $N_\data$ is the total number of SFs occupied by UL or DL data TBs and $n_\rep$ is the number of repetitions used for each TB, i.e., the number of times each TB is repeated including its first transmission. Note that the TB repetitions introduce redundancy to improve the error rates, and are not the same as HARQ re-transmissions. The value of $n_\rep$ is chosen based on the adaptive MCS used by the UE. HARQ re-transmissions, on the other hand, are repeated transmissions of one or more repetitions of a TB when the TB is not successfully acknowledged. A detailed explanation of the HARQ operation can be found extensively in the literature, e.g.,~\cite[Ch. 10]{sesia2011lte}.

Eq.~\eqref{eq:suf} shows that SUF indicates the proportion of time spent on transmitting the payload as opposed to transmission overheads. We therefore compute the useful data rate, $R$, as 
\begin{equation}\label{eq:throughput_general}
    R = \eta_\suf \frac{n_\tb}{t_\tb},
\end{equation}
where $n_\tb$ and $t_\tb$ are the TBS in bits and time spanned by one TB, respectively. Therefore, for a given TBS corresponding to the adaptive MCS chosen, we can maximize the data rate by maximizing the SUF, i.e., by reducing the transmission overheads. Since higher values of RTT can increase $N_\hc$, it can be seen from~\eqref{eq:suf} that it thus deteriorates SUF and the useful data rate.

\subsection{\textcolor{black}{Related Work}} \label{sec:prior_art}
IoT-NTN studies have thus far, rightly, focused predominantly on initial access aspects, such as analysis of link budget~\cite{conti2020nb, gineste2017narrowband}, challenges of Doppler effects and solutions to counter them~\cite{charbit2020space, conti2020doppler, gineste2017narrowband}, and issues related to random access~\cite{kodheli2020random, chougrani2021nb, gineste2017narrowband, 9502071}.
Enhancing scheduling for IoT-NTN networks has also been considered in the past, but from a system-level perspective, i.e., enhancements at the BS-end for scheduling transmissions among users~\cite{kodheli2018resource, kodheli2019uplink}. The solutions proposed are beneficial in countering the impact of increased Doppler shift in IoT-NTN systems, but do not address the link-level packet scheduling issue that we consider in this work to enhance data rates. 

\textcolor{black}{In the context of NTN, solutions that have previously been proposed to counter the impact of large propagation delay can be roughly classified into three categories:
\begin{enumerate}[A)]
    \item targeting a low initial block error rate (iBLER),
    \item disabling the use of HARQ feedback at the medium access control (MAC) layer, and
    \item increasing the number of HARQ processes.
\end{enumerate}
In the following, we consider solutions from each of these categories and discuss the reasons why they are not applicable to solve the problem at hand. We also provide qualitative comparisons to demonstrate how our proposed solutions are more suitable than those in the existing art. \textcolor{black}{Additionally, we also compare HARQ techniques to other automatic repeat request (ARQ) methods that have been proposed in the past.}}

\subsubsection{Low Target iBLER}\label{subsec:low_ibler}
\textcolor{black}{The issue of stop-and-wait gaps produced due to the lengthy RTT is recognized in~\cite{giordani2020non, barbau2020nb} for GEO systems. These works acknowledge that packet errors and the resulting re-transmissions can significantly increase transmission gaps especially for high altitude GEO links and thereby reduce the data rate.} While~\cite{giordani2020non} only discusses the open issue of stop-and-wait gaps,~\cite{barbau2020nb} proposes a solution to counter this by simply improving robustness, i.e., targeting a low iBLER to avoid re-transmissions. \textcolor{black}{Although using low iBLERs can help in reducing re-transmissions, previous simulation studies have shown that a high target iBLER instead is optimal to reduce the overall UE power consumption. This is because a larger target iBLER demands fewer repetitions and can accommodate a higher coding rate~\cite{R1_1705013}.} The study in~\cite{R1_1705013} however only considers terrestrial networks operating with negligible propagation delay. When a large iBLER is used in NTNs, it does result in increased stop-and-wait gaps. Therefore, our solution, aimed precisely at reducing these stop-and-wait gaps with smart scheduling, presents a method to use a high iBLER for battery life optimization for IoT-NTN devices, while at the same time also increasing UE battery life by enabling a higher target iBLER. \textcolor{black}{Nevertheless, note that our proposed solution is also compatible with designs that use a sub-optimal low iBLER transmission, but with reduced throughput gains.}

\subsubsection{Disabled HARQ}
Although link-level scheduling research in IoT-NTN is still in its infancy, the issue has been studied extensively for supporting eMBB applications in NR-NTN. An overview of these methods is reported in~\cite{tr_38821}. We evaluate if the methods suggested for NR-NTN can be adapted also for IoT-NTN. One solution to solve the issue of stop-and-wait gaps in HARQ transmissions is to disable the use of HARQ feedback~\cite[Sec. 6]{tr_38821},~\cite{lin20215g}. In this case, re-transmissions can be handled at the radio link control (RLC) layer. However, several problems exist with this technique. First, the benefits of HARQ combining (e.g., Chase combining or incremental redundancy) at the physical (PHY) or MAC layer are absent, and thus, more re-transmissions may be needed, which then results in poor overall spectral efficiency. Furthermore, relying on RLC layer for re-transmission introduces additional user-plane latency and increased jitter when errors occur due to the high re-transmission timeout. This is because, without HARQ feedback, the network relies on the RLC feedback. In this case, the status report, which contains the ACK or negative ACK (NACK) feedback, can be severely delayed. \textcolor{black}{For example, when there is no UL data to transmit, the UE must send a scheduling request on PUCCH to request the scheduling of UL data just to send the status report~\cite{ts_36321}. This can take longer than three times the RTT, depending on PUCCH configuration.} Furthermore, when there are burst errors, the status report can only be sent after a valid packet is received, which also causes increased latency. Additionally, if the last physical data unit is in error, the status report is not sent until the transmitter requests it. \textcolor{black}{A possible solution to counter the delayed status reporting is to let the transmitter request an RLC status report often via the \textit{poll bit}~\cite{ts_36331}. While this reduces signaling latency, it introduces a large signaling overhead, which may be tolerable for eMBB applications but is significant for IoT-NTN scenarios. }

\subsubsection{Increasing the Number of HARQ Processes}
An alternative solution to reduce or eliminate the stop-and-wait gaps is to increase the number of HARQ processes, $N_\harq$. \textcolor{black}{Increasing $N_\harq$ is associated with two major issues. First, it requires a larger soft buffer size at the receiver. This introduces cost and complexity overheads, and is not suitable for low-cost IoT-NTN devices. Second, implementing a higher number of HARQ processes for HD-FDD IoT-NTN UEs using state-of-the-art scheduling methods requires the use of MTBG and ACK bundling. MTBG and ACK bundling increases the system complexity and may not be implemented in all networks, especially those catering to low-cost IoT devices. Furthermore, ACK bundling is only efficient when the iBLER is small. But as discussed earlier in Section~\ref{subsec:low_ibler}, a lower iBLER results in poorer spectral efficiency, lower speed, and higher UE power consumption.  Therefore, increasing the $N_\harq$ with the existing scheduling methods to fill the stop-and-wait gaps is not ideal for IoT-NTN systems. Nevertheless, we show in the following sections that our proposed technique for scheduling IoT NTN TBs not only improves spectral efficiency without necessarily requiring an increase in $N_\harq$, but also allows the usage of an arbitrary number of parallel HARQ processes. This thereby also allows for increasing $N_\harq$ in the uplink without incurring the two issues described above.}

\subsubsection{\textcolor{black}{Use of Alternative ARQ Techniques}}
\textcolor{black}{Fig.~\ref{fig:legacy_timing} shows an example of the use of one HARQ process, which is similar to a stop-and-wait ARQ method. As discussed before, such a scheduling results in large stop-and-wait delays, especially in lengthy propagation time environments, such as NTN. Other ARQ methods, such as go-back-N (GBN) ARQ and selective repeat (SR) ARQ are known to provide better net throughput than stop-and-wait ARQ, with SR-ARQ outperforming GBN-ARQ~\cite[Ch. 5]{leon2000communication}. The parallel HARQ processes architecture used in LTE-M and NB-IoT uses techniques identical to SR-ARQ for handling retransmissions, together with incremental redundancy for soft-combining~\cite{beh2007performance},~\cite[Ch. 10]{sesia2011lte}, for increasing transmission robustness. Comparative performance evaluations have been performed in the past to show that error detection and correction using HARQ outperforms the ARQ methods~\cite{rajhi2020comparison}. Therefore, in this paper, we focus on developing a HARQ-based scheduling strategy, which is also compatible with standardized IoT-NTN designs. This enables our solution to be readily implemented on commercial cellular IoT-NTN UEs.    }

With this backdrop, we propose our flexible HARQ scheduling solutions in the following section, that do not require disabling HARQ processes or demand any \textit{mandatory} increase in $N_\harq$. 

\vspace{.2in}

\section{Proposed Solutions}\label{sec:proposed_solutions}
The value of $N_\harq$ required to fill up all the stop-and-wait gaps in its entirety is
\begin{equation}
    N_\harq \geq \frac{\tau_\rtt}{t_\tb},
\end{equation}
where $\tau_\rtt$ is the RTT in seconds. From Table~\ref{table:rtt}, we can observe that $N_\harq$ must be increased to up to $42$ under typical operating conditions where one TB spans $1$ SF. The current LTE-M and NB-IoT specifications only allow a maximum of $8$ and $2$ HARQs, respectively~\cite{ts_36213, ts_36321}. This increase in $N_\harq$ is impractical for low cost mMTC and C-IoT UEs, where a high $N_\harq$ introduces increased complexity due to the required size of the soft-buffer. Furthermore, it also demands corresponding increments in the size of the downlink control information (DCI) bits. 

For the case of IoT-NTN, which uses few physical resource blocks (PRBs) and operates under high path loss environments, multiple repetitions of the TB are often required to achieve reliable communication. Therefore, the required $N_\harq$ is
\begin{equation}
    N_\harq \geq \frac{\tau_\rtt}{n_{\rep,\data} t_\tb},
\end{equation}
where $n_{\rep,\data}$ is the number of data block repetitions, i.e., $n_{\rep, \data} = n_{\rep, \pdsch}$ or $n_{\rep, \data} = n_{\rep, \pusch}$, for indicating repetitions on the PDSCH or the PUSCH, respectively.
In Section~\ref{sec:results}, we present the results from a comprehensive link-level simulation evaluation to show that IoT-NTN does not require any further increase in $N_\harq$ for MTC applications, and up to $N_\harq = 4$ for NB-IoT applications to achieve reliable communications while also filling the stop-and-wait gaps. 

\begin{figure*}[t]
	\centering
	\includegraphics[width=15cm]{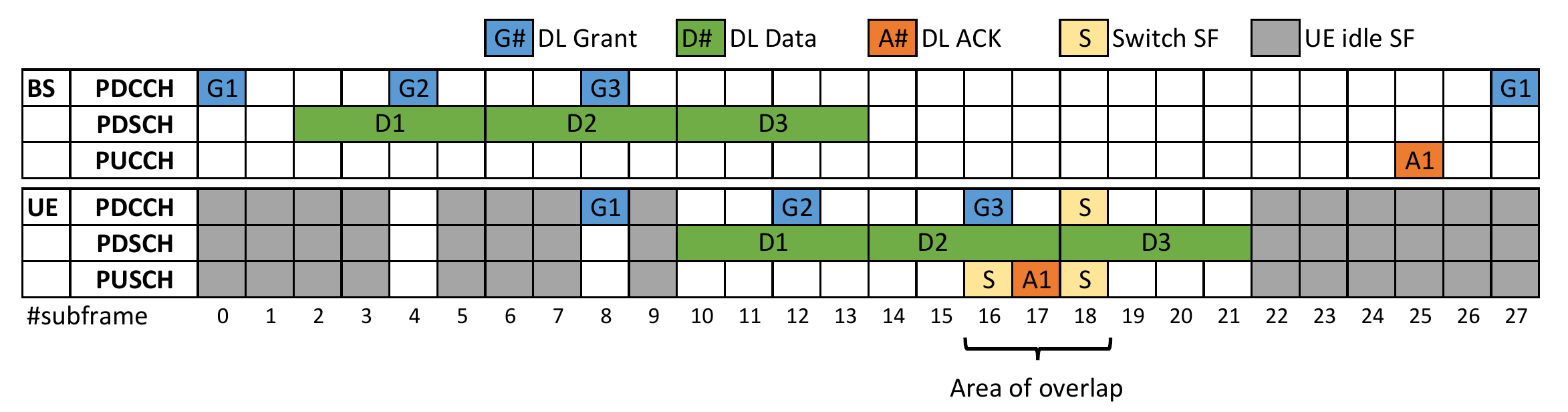} 
	\caption{\textcolor{black}{An illustration of the scheduling conflict with TB overlap using the current-day method.}}
	\label{fig:tb_overlap}
\end{figure*}

The issue with the current HARQ implementation is that the delay between the DL data and the corresponding ACK, $n_{\text{DD2A}}$, and similarly in the uplink, the delay between the UL grant and UL data, $n_{\text{UG2D}}$, are fixed. Therefore, 
\begin{align}
    i_{\ack,j} = i_{\text{DLdata},j} + n_{\text{DD2A}} +1, \;\; & \text{for DL} \\
    i_{\text{ULdata},j} = i_{\text{UG},j} + n_{\text{UG2D}} + 1, \;\; & \text{for UL},
\end{align}
where $i_{\ack,j}$, $i_{\text{DLdata},j}$, $i_{\text{ULdata},j}$, and $i_{\text{UG},j}$ are the positions in SFs of the ACK for the $j$th TB, the last SF of the $j$th DL TB, the first SF of the $j$th UL TB, and the last SF of the $j$th UG, respectively. When $n_\rep > n_{\text{DD2A}}$, $i_{\ack,j}$ overlaps with $i_{\text{DLdata},k}$ for $k>j$ or the switching SF. Similarly, when $n_\rep > n_{\text{UG2D}}$, $i_{\text{DLdata},k}$ for $k>j$ overlaps with $i_{\text{ULdata},j}$. \textcolor{black}{We show an example of this issue in the downlink in Fig.~\ref{fig:tb_overlap}, with $n_\rep = 4$ and $n_{\text{DD2A}} = 3$. We notice that $i_{\ack,1}$ on the PUCCH overlaps with $i_{\text{DLdata},2}$ on the PDSCH, which is inoperable for HD-FDD UEs.}

Thus, the current HARQ configuration can only support one TB per HARQ cycle when $n_\rep > n_{\text{DD2A}}, n_{\text{UG2D}}$. We can re-write~\eqref{eq:suf} under such an operation as 
\begin{align}\nonumber
    \eta_{\suf, \text{DL}} = &(n_{\rep, \pdcch}+n_{\rep, \pdsch}+n_{\rep, \pucch}\\ &+ N_{\text{DG2D}} + n_{\text{DD2A}} + N_\switch)^{-1}
\end{align}
and
\begin{align}
    \nonumber
    \eta_{\suf, \text{UL}} = &(n_{\rep, \pdcch}+n_{\rep, \pusch}\\ &+ n_{\text{UG2D}} + N_\switch)^{-1},
\end{align}
for DL and UL, respectively, where $n_{\rep, \pdcch}$ and $n_{\rep, \pucch}$ represent the number of repetitions of grant on PDCCH and ACK on PUCCH, respectively, $N_{\text{DG2D}}$, $n_{\text{DD2A}}$, and $n_{\text{UG2D}}$ and the processing delays between DL grant and DL data, DL data and ACK, and UL grant and UL data, respectively, and $N_\switch$ is the switching delay, all in units of SFs. 

We show in Section~\ref{sec:results} that $n_{\rep, \pusch} > n_{\text{DD2A}}, n_{\text{UG2D}}$ always holds true, especially with the delay values specified in the legacy LTE-M protocol. Therefore, irrespective of the $N_\harq$ that can be used, only one TB can be scheduled per HARQ cycle. We refer back to the example of NTN DL HARQ transmissions with fixed DD2A and a time of flight of $8$~SF units in Fig.~\ref{fig:ntn_mtc_harq}. In this example, we use a fixed DD2A of $3$~SFs, $4$~SFs of data repetitions, and $1$~HARQ process. We notice that the processing delays and the time of flight between the UE and BS result in several idle SFs in the UE due to the single TB in one HARQ cycle. When multiple TBs are scheduled on one HARQ cycle, the burden of additional delays can be spread over those multiple TBs. This reduces the overall overheads per TB and improves the SUF. It is even more beneficial in the context of NTN, since the over-the-air (OTA) travel time can be exploited to overlap the bidirectional UL and DL signals in air. To this end, we propose using flexible values of $n_{\text{DD2A}}$ and $n_{\text{UG2D}}$ to accommodate multiple TBs per HARQ cycle.

\textcolor{black}{Our proposal of using TB-specific DD2A and UG2D delays\footnote{\textcolor{black}{DD2A delay can also be considered as a PDSCH to PUCCH delay, and UG2D delay can also be considered as a PDCCH to PUSCH delay.}}, i.e., $n_{\text{DD2A}, j}$ and $n_{\text{UG2D},j}$, eliminates the overlap between the data TBs and/or the switching SFs.} Thereby, we ensure that an arbitrary number of TBs can be scheduled within one HARQ cycle. The SUF in this case can be expressed as in~\eqref{eq:suf_dl} and~\eqref{eq:suf_ul}, respectively, where $N_{\text{TBPHC}}$ is the number of TBs scheduled in one HARQ cycle and $n_{\text{DD2A, min}}$ and $n_{\text{UG2D, min}}$ are the minimum mandatory DD2A and UG2D delays that must be used to ensure that sufficient processing time is available at the UE to process DL data and UL grants, respectively.
\begin{figure*}[b]  
	\begin{align}\label{eq:suf_dl}
		\eta_{\suf, \text{DL}} &= \frac{N_{\text{TBPHC}}}{n_{\rep,\pdcch} + N_{\text{DG2D}} +  N_{\text{TBPHC}}n_{\rep, \pdsch} + n_{\rep, \pucch} + \max(n_{\text{DD2A, min}}, (N_{\text{TBPHC}}-1)n_{\rep, \pucch}) + 2N_\switch}\\ \label{eq:suf_ul}
		\eta_{\suf, \text{UL}} &= \frac{N_{\text{TBPHC}}}{n_{\rep,\pdcch} + \max(n_{\text{UG2D, min}}, (N_{\text{TBPHC}}-1)n_{\rep,\pdcch}) + N_{\text{TBPHC}}n_{\rep, \pdsch} + 2N_\switch}
	\end{align}  
\end{figure*}
While any $N_{\text{TBPHC}} > 1$ can be chosen to obtain higher $\eta_{\suf, \text{DL}}$ and $\eta_{\suf, \text{UL}}$, the limit on the maximum $N_{\text{TBPHC}}$ that can be scheduled is set by the maximum $N_\harq$ supported by the specification. The relation between $N_{\text{TBPHC}}$ and $N_\harq$ is dependent on the RTT of the network and can be expressed as in~\eqref{eq:num_harq}, where $N_{\text{A2G}}$ is the ACK processing delay at the BS before scheduling the HARQ process whose TB is acknowledged. 
\begin{figure*}[t]  
	\begin{align}\label{eq:num_harq}
		&N_\harq = \left \lceil N_{\text{TBPHC}} \left( 1+   \frac{\tau_\rtt + N_{\text{A2G}}}{t_\tb (n_{\rep, \pdcch} + N_{\text{DG2D}} + N_{\text{TBPHC}}(n_{\rep, \pdsch} + n_{\rep, \pucch})  + 2N_\switch  )} \right) \right \rceil
	\end{align}  
\end{figure*}

\textcolor{black}{Next, we present the methods to compute, choose, and signal the variable delays from the BS to the UE.} Toward this end, we consider three different scenarios of HARQ scheduling. \textcolor{black}{The use of MTBG and ACK bundling may not be considered together due to their complexity, and has also already been investigated for delay flexibility in terrestrial networks~\cite{US20210028918} (and can therefore be extended to NTNs if needed). Therefore, we consider the three other cases of \textit{grouping} grants and ACKs, namely, MTBG without ACK bundling, STBG without ACK bundling, and STBG with ACK bundling. }

\subsubsection{No ACK Bundling}
\begin{figure}[t]
	\centering
	\includegraphics[width=8.5cm]{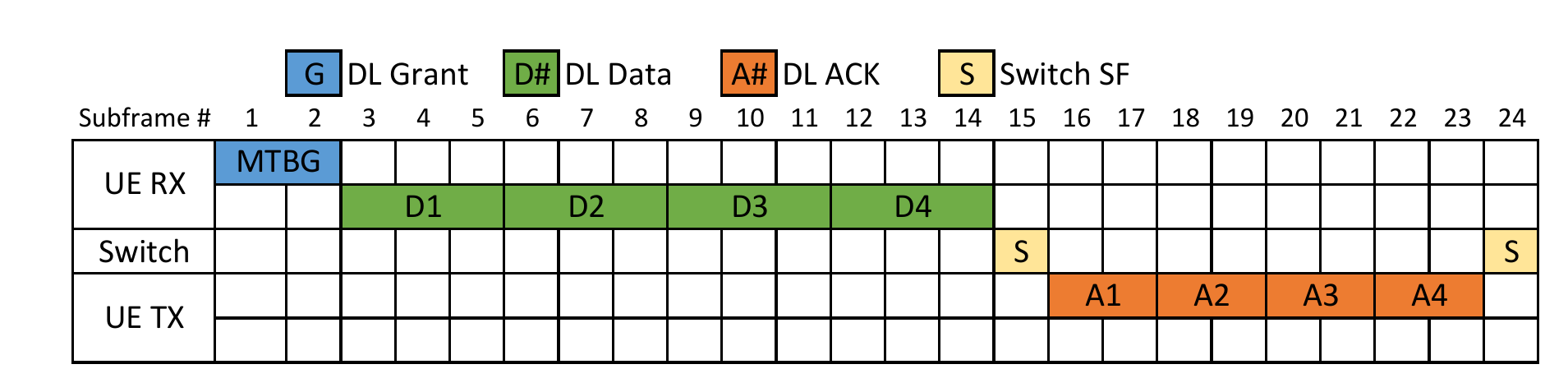} 
	\caption{Timing diagram for DL operation with MTBG.}
	\label{fig:mtbg}
\end{figure}
An example of MTBG without ACK bundling for DL transmission is shown in Fig.~\ref{fig:mtbg}. When STBG is used instead, the grants are split to schedule each TB individually. The variable DD2A of the $j$th TB is the sum of the time left for the remaining TBs scheduled in the MTBG, the ACKs corresponding to all TBs from $1, 2, ... j-1$, and the switching SFs. Therefore,
\begin{equation}\label{eq:dd2a_mtbg}
	n_{\text{DD2A},j} = (N_{\text{TBPHC}}-j)n_{\rep, \pdsch} + (j-1)n_{\rep, \pucch} + N_\switch.
\end{equation}
All the parameters used in~\eqref{eq:dd2a_mtbg} are already available to the UE to compute $n_{\text{DD2A},j}$ since $ N_{\text{TBPHC}}$, $n_{\rep, \pdsch}$, and $j$ are extracted from the MTBG and $n_{\rep, \pucch}$ is a radio resource control (RRC) configured parameter. Therefore, the UE requires no additional signaling from the BS to compute $n_{\text{DD2A},j}$. However, with the use of STBG, $n_{\rep, \pdsch}$ is conveyed by legacy grants, while the identifier $j$ is to be signaled by the BS when scheduling the $j$th TB. Similarly, $N_{\text{TBPHC}}$ can also be explicitly signaled to the UE if the BS chooses to configure an $N_{\text{TBPHC}}$ corresponding to an $N_\harq$ that is lower than the supported maximum value.
The same principle can also be applied in the UL to obtain the variable UG2D delay as 
\begin{equation}\label{eq:ug2d_mtbg}
	n_{\text{UG2D},j} = (N_{\text{TBPHC}}-j)n_{\rep, \pdcch} +(j-1)n_{\rep,\pusch}+N_\switch.
\end{equation}
Similar to the case of DL, all parameters required to compute~\eqref{eq:ug2d_mtbg} is acquired by the UE as in the case of DL transmission. 

\begin{figure}[t]
	\centering
	\includegraphics[width=8.5cm]{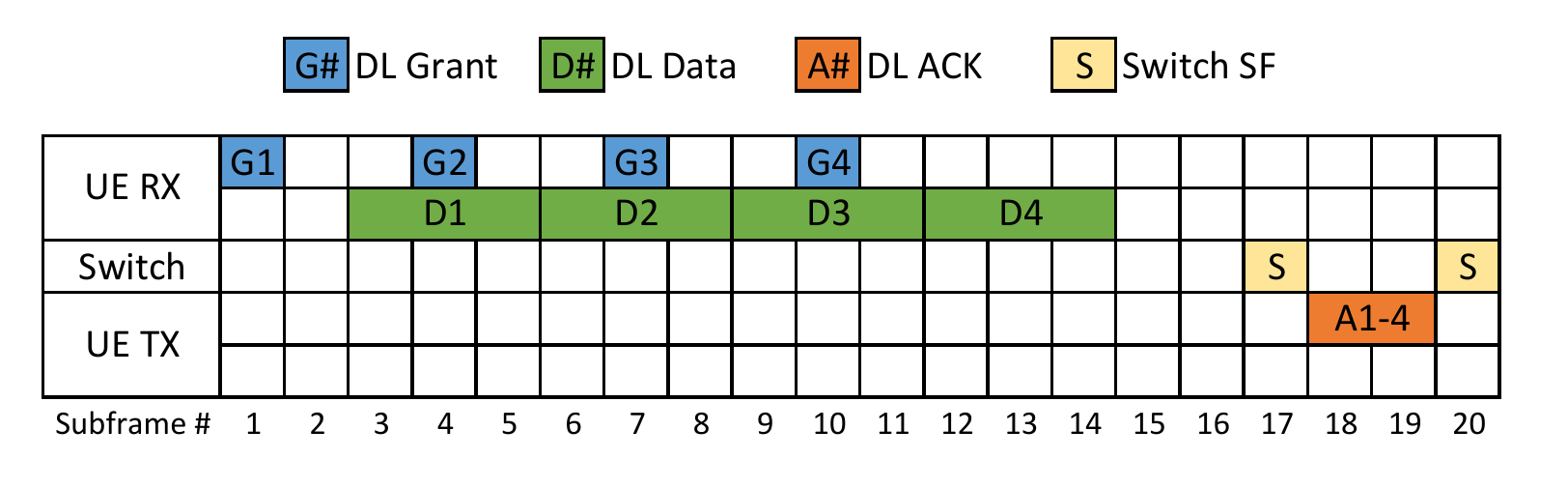} 
	\caption{Timing diagram for DL operation with ACK bundling.}
	\label{fig:ack_bundle}
\end{figure}

\subsubsection{With ACK Bundling}
The condition of ACK bundling is only applicable in the DL since there is no notion of acknowledgment in the UL. We show an example timing diagram in Fig.~\ref{fig:ack_bundle}, where the acknowledgment for four TBs are bundled together into two TTIs. The number of TTIs to bundle the plurality of ACKs may be up to the network implementation. When the ACKs are bundled into a single SF,~\eqref{eq:dd2a_mtbg} can be modified as 
\begin{align}\nonumber
	n_{\text{DD2A},j}=(N_{\text{TBPHC}}-j)n_{\rep, \pdsch} + & \left \lfloor \frac{j-1}{n_{\text{bundle}}} \right \rfloor  n_{\rep, \pucch} \\ \label{eq:ackbundle} &+ N_\switch,
\end{align}
\textcolor{black}{where $n_{\text{bundle}}$ is the number of ACKs that are bundled within one TTI. $n_{\text{bundle}}$ can be learned by the UE during RRC configuration and hence requires no signaling overhead. }

The above analyses assume that the number of repetitions used by all TBs within a HARQ cycle is the same, as is the case in current LTE-M and NB-IoT operations. However, the BS may choose to use different MCS for each TB based on the type of data being transmitted. While this method is currently not supported in LTE-M or NB-IoT specifications, we address this condition to make our solution future-proof when such an adaptive transmission technique may be implemented in the future. In this case, $n_{\rep, \pdsch}$ and $n_{\rep, \pusch}$ of each of the remaining or previous TBs are required to compute the DD2A and UG2D delays, respectively. Therefore,~\eqref{eq:dd2a_mtbg}, \eqref{eq:ug2d_mtbg}, and~\eqref{eq:ackbundle} can be modified as
\begin{align}
	n_{\text{DD2A},j} = &\sum \limits_{k=j}^{N_{\text{TBPHC}}} n_{\rep, \pdsch, k} + (j-1)n_{\rep, \pucch} + N_\switch, \\ \nonumber
	n_{\text{UG2D},j} = &(N_{\text{TBPHC}}-j)n_{\rep, \pdcch} + \sum \limits_{k=1}^{j-1} n_{\rep,\pusch,k} \\ &+N_\switch,\\ \nonumber
	n_{\text{DD2A},j} = &\sum \limits_{k=j}^{N_{\text{TBPHC}}} n_{\rep, \pdsch, k} + \left \lfloor \frac{j-1}{n_{\text{bundle}}} \right \rfloor  n_{\rep, \pucch} \\ \label{eq:dd2a_bundle_var_rep}
					  & + N_\switch,
\end{align}
respectively, where $n_{\rep, \pdsch, k}$ and $n_{\rep,\pusch,k}$ are the number of repeats on the PDSCH and PUSCH for the $k$th TB, respectively. 

\section{Results}\label{sec:results}
In this section, we present numerical results of the increase in data rates achieved with the use of our proposed solutions. The primary reason for the throughput gains obtained by our method can be attributed to the higher number of TBs that can be supported in one HARQ cycle by our variable delay design. \textcolor{black}{However, we recognize that increasing the number of TBs in one HARQ cycle and the consequent possible use of an increased $N_\harq$ (for the case of NB-IoT) specifically to fill the entire RTT results in a higher complexity at the receiver due to a larger size of the soft-buffer.} \textcolor{black}{This condition is typically not preferable when the receiver is the UE (i.e., for DL communications). On the other hand, a higher receiver complexity introduced by the increased buffer size is negligible in the uplink where the BS receives the TBs.} Hence, considering practical implementation scenarios, we focus on the uplink to evaluate the throughput gains achievable using our proposed methods. \textcolor{black}{Note that an increase in $N_\harq$ is not mandatory for the operation of our proposed methods, and performance improvements using our solution can be achieved regardless of the increase. }

From~\eqref{eq:suf_dl}$-$\eqref{eq:dd2a_bundle_var_rep}, we notice that the throughput achievable with the use of our method relies on the number of data TB  repetitions. To this end, we first begin our evaluation campaign with a link-level simulation of a point-to-point IoT-NTN uplink path to determine a suitable value of $n_{\rep, \pusch}$ required to achieve a target BLER.

\subsection{Simulation Settings}
We derive a majority of our simulation settings from the relevant 3GPP technical reports and technical documents related to NTN~\cite{tr_38821, tr_38811, zte_ntn}. As suggested by 3GPP, we consider the NTN UEs to be enabled with a global navigation satellite system (GNSS) ability such that it can perform pre- and post-compensation of the frequency offset~\cite{ran1_103e}. Therefore, we apply a maximum residual frequency offset of $34$~Hz after post-compensation at the satellite and pre-compensation along with continuous frequency tracking at the UE~\cite{ericsson_ntn}. We use the 3GPP recommended NTN tapped delay line (TDL) non-line-of-sight (NLOS) channel model to investigate the performance of our solution. We scale the power delay parameters of the reference TDL-A model~\cite{tr_38811} according to the desired value of delay spread specified for suburban environment~\cite{tr_38821, tr_38901}. We choose two different values of $n_\tb = \{144, 504 \}$ to investigate the impact of TBS on BLER and spectral efficiency.  We list all simulation parameters in Table~\ref{table:evaluation_settings}.

\begin{table}[]
\centering
\caption{Evaluation Settings}\label{table:evaluation_settings}
\begin{tabular}{c|c}
\hline
Parameter        & Value         \\ \hline\hline
$n_\tb$                                  & 144, 504 bits \\ \hline
Modulation                           & QPSK          \\ \hline
No. of PRBs                       & 1             \\ \hline
SFs for channel estimation          & 5             \\ \hline
No. of HARQ processes             & 1             \\ \hline
Channel model                        & NTN TDL-A \\ \hline
UE environment                          & Suburban NLOS \\ \hline
No. of transmit antenna           & 1             \\ \hline
No. of receive antenna            & 2             \\ \hline
UE Speed                             & $15$~km/h       \\ \hline
UE-satellite elevation angle            & $30$~degrees      \\ \hline
Feeder-satellite elevation angle            & $10$~degrees      \\ \hline
Residual frequency offset            & Uniformly distributed in \\                                           & [$-34$~Hz, $34$~Hz]         \\ \hline
$f$                             & $2$~GHz         \\ \hline
Target BLER                      & $10\%$ \\ \hline
$P_{\text{EIRP}}$               & $23$~dBm \\ \hline
$G/T$                           & $-4.9$~dB/T \\ \hline
$\delta$                        & $180$~kHz \\ \hline
$\beta_{\text{atm}}$            & $0.07$~dB \\ \hline
$\beta_{\text{shadow}}$         & $3$~dB \\ \hline
$\beta_{\text{scint}}$          & $2.2$~dB \\ \hline
$\beta_{\text{polar}}$          & $0$~dB \\ \hline

\end{tabular}
\end{table}

\subsection{Numerical Results}
\begin{figure}[t]	
	\begin{center}
		\includegraphics[clip,width=0.95\columnwidth]{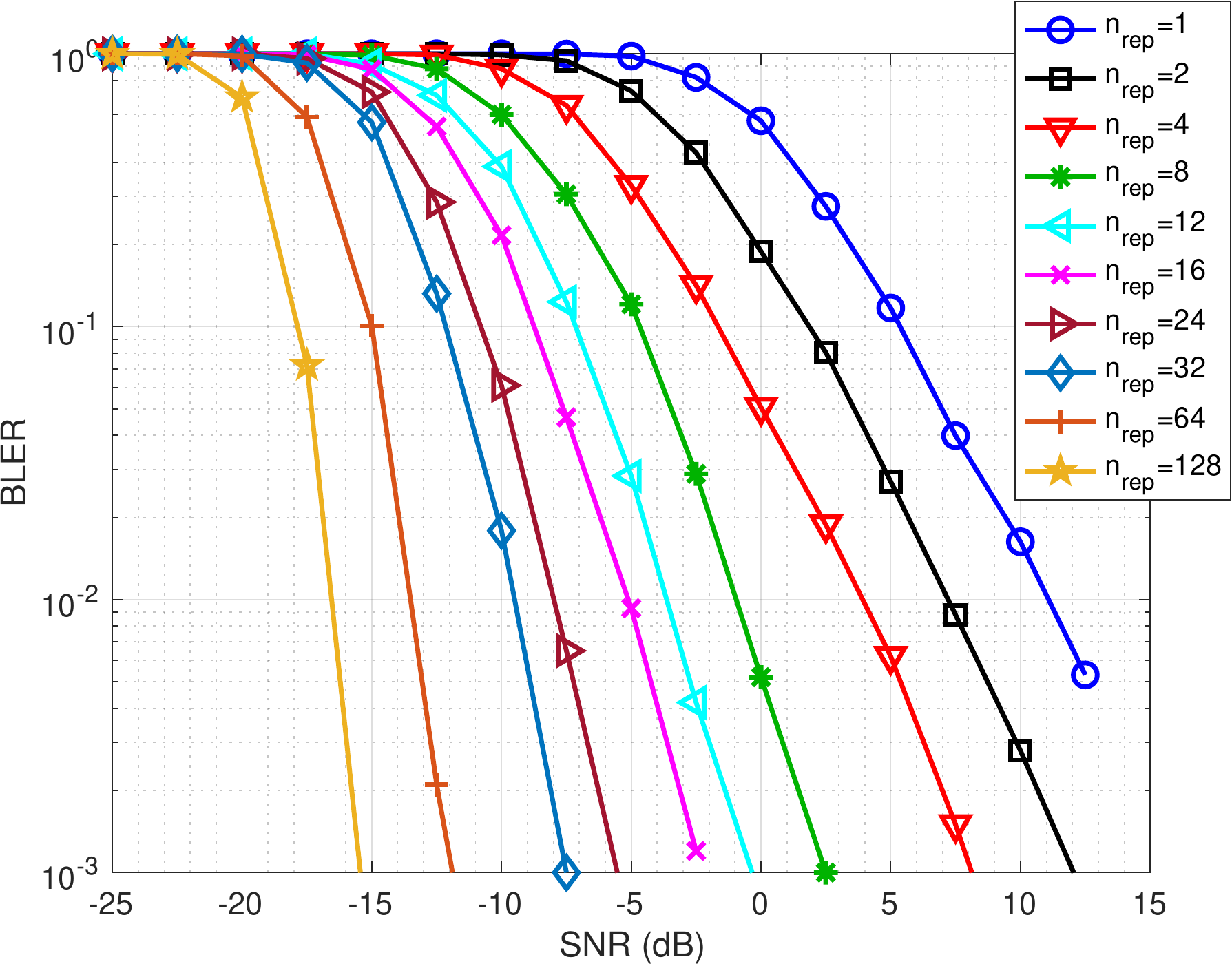}
		\caption{BLER vs SNR for different repetitions ($n_\text{rep}$) of PUSCH with TBS of 144 bits in NTN TDL-A channel.}
		\label{fig:BLERresultsTDLA144}
	\end{center}
\end{figure}
\begin{figure}[t]	
	\begin{center}
		\includegraphics[clip,width=0.95\columnwidth]{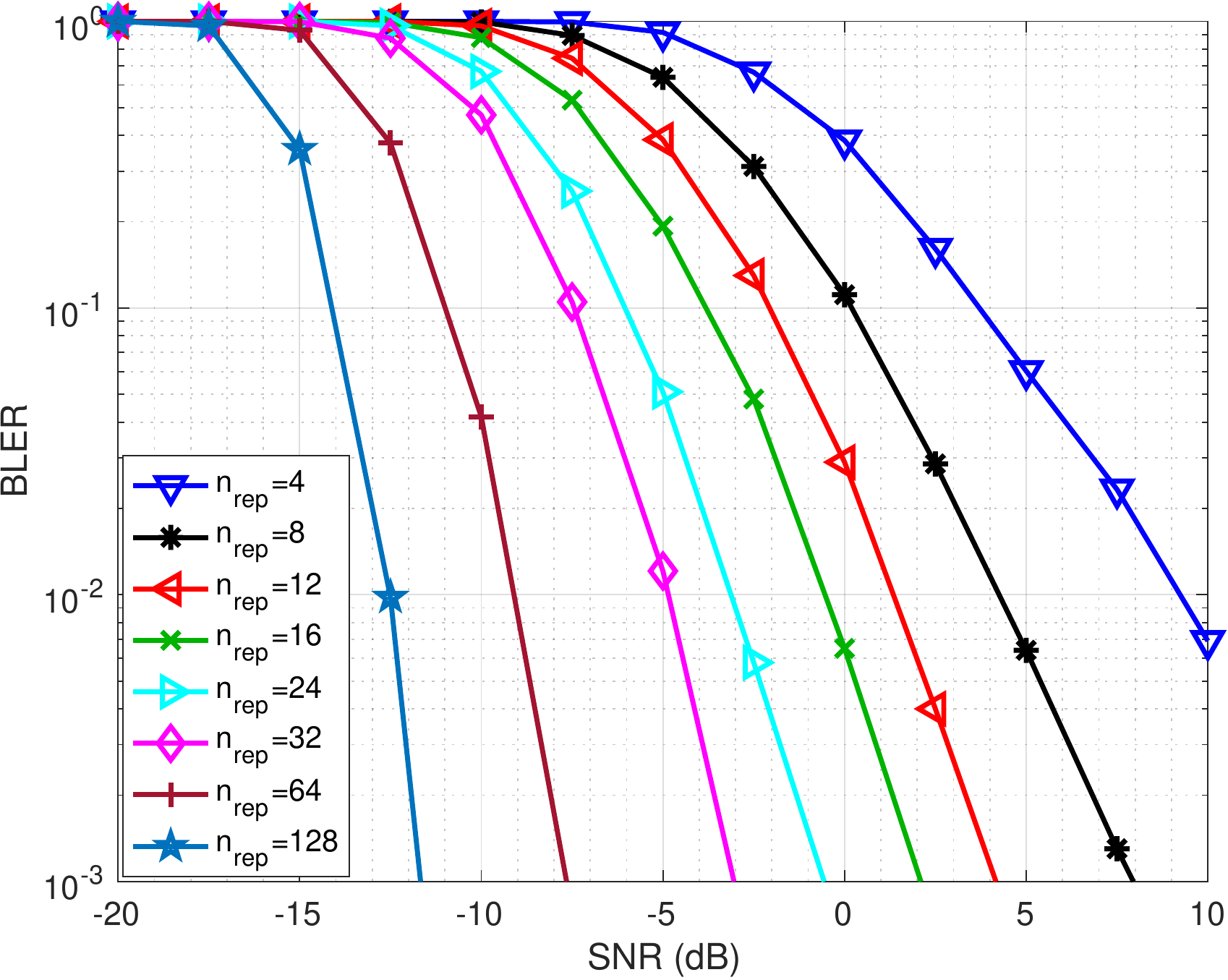}
		\caption{BLER vs SNR for different repetitions ($n_\text{rep}$) of PUSCH with TBS of 504 bits in NTN TDL-A channel.}
		\label{fig:BLERresultsTDLA504}
	\end{center}
\end{figure}

We begin by presenting our link-level simulation results. 

\subsubsection{Number of Repetitions}
In Figs.~\ref{fig:BLERresultsTDLA144} and~\ref{fig:BLERresultsTDLA504}, we present the results of the variation of BLER for different possible signal-to-noise ratio (SNR) values for TDL-A channel. As expected, the BLER achieved with increased number of repetitions is lower due to the increased redundancy. \textcolor{black}{Based on our desired operating conditions (i.e., SNRs), we use the results in Figs.~\ref{fig:BLERresultsTDLA144} and~\ref{fig:BLERresultsTDLA504} to choose the required number of repetitions to achieve a target BLER. We then use these numbers to present the throughput gains of our proposed solutions in the following.}

\subsubsection{Spectral Efficiency and Throughput Gain} \label{subsec:throughput_gain}
We focus on the UL throughput gain provided by the use of our variable delay methods using both LTE-M and NB-IoT based UEs. We consider the two LEO satellite access types of LEO600 and LEO1200 that consist of satellites at altitudes of $600$~km and $1200$~km, respectively. We use an elevation angle of $30$~degrees, which provides $\tau_\rtt = 20$~ms and $\tau_\rtt = 34$~ms for LEO600 and LEO1200 scenarios, respectively. Next, we compute the operating SNR condition to determine a suitable value of $n_\rep$ to use for evaluating our proposed solutions. We compute the SNR, $\gamma$, as 
\begin{equation}
    \gamma = \frac{P_{\text{EIRP}} G }{kT \beta_{\text{PL}} \beta_{\text{atm}} \beta_{\text{shadow}} \beta_{\text{scint}} \beta_{\text{polar}} \delta},
\end{equation}
where $\beta_{\text{PL}}$ is the free space path loss given by~\cite{tr_38811}
\begin{equation}
    \beta_{\text{PL}} = 10^{3.245 + \log_{10}(f^2) + \log_{10}(d^2)},
\end{equation}
$\delta$ is the signal bandwidth, $P_{\text{EIRP}}$ is the effective isotropically radiated power from the NTN UE, $G/T$ is the antenna-gain-to-noise-temperature value of the satellite antenna, $k$ is the Boltzmann constant, $f$ is the carrier frequency in GHz, $d$ is the distance between the UE and the satellite, and $\beta_{\text{atm}}$, $\beta_{\text{shadow}}$, $\beta_{\text{scint}}$, $\beta_{\text{polar}}$ are the atmospheric loss, shadow fading margin, scintillation loss, and polarization loss, respectively. We compute $d$ based on the satellite altitude and the elevation angle listed in Table~\ref{table:evaluation_settings}. Using the values from Table~\ref{table:evaluation_settings}, we obtain $\gamma = -0.2$~dB and $\gamma = -5.6$~dB for LEO600 and LEO1200 scenarios, respectively. These conditions are also consistent with the suggested link budget evaluations presented for NR-NTN in~\cite{tr_38821}. 

We find from Figs.~\ref{fig:BLERresultsTDLA144} and~\ref{fig:BLERresultsTDLA504} that for a fixed target BLER, choosing a higher value of $n_\tb$ provides better spectral efficiency. For example, for a target BLER of $10\%$ at the operating value of $\gamma = -5.6$~dB for LEO1200, $n_{\rep, \pusch} = 12$ and $n_{\rep, \pusch} = 24$ for $n_\tb = 144$ and $n_\tb = 504$, respectively. This results in a spectral efficiency of $12$ and $21$ bits/PRB for $n_\tb = 144$ and $n_\tb = 504$, respectively. This phenomenon of higher spectral efficiency for larger TBS is also true across satellite access types and target BLERs. Therefore, we perform our throughput gain evaluation for our proposed method with $n_\tb = 504$. We extract the corresponding numbers for $n_\rep$ for both satellite access types at their operating SNR values from Fig.~\ref{fig:BLERresultsTDLA504} as $n_\rep = 12$ and $n_\rep = 24$ for LEO600 and LEO1200 scenarios, respectively. \textcolor{black}{This also clearly shows that $n_{\rep, \pusch}$ for both LEO600 and LEO1200 cases is greater than $n_{\text{DD2A}}$ and $n_{\text{UG2D}}$. As demonstrated in Section~\ref{sec:proposed_solutions}, this condition results in the PUCCH SFs overlapping the PDSCH time slots in the DL and PDCCH SFs overlapping with PUSCH in the UL, respectively, for the case of fixed DD2A and UG2D methods of the state-of-the-art.}

\begin{figure}[t]
	\centering
	\includegraphics[width=7cm]{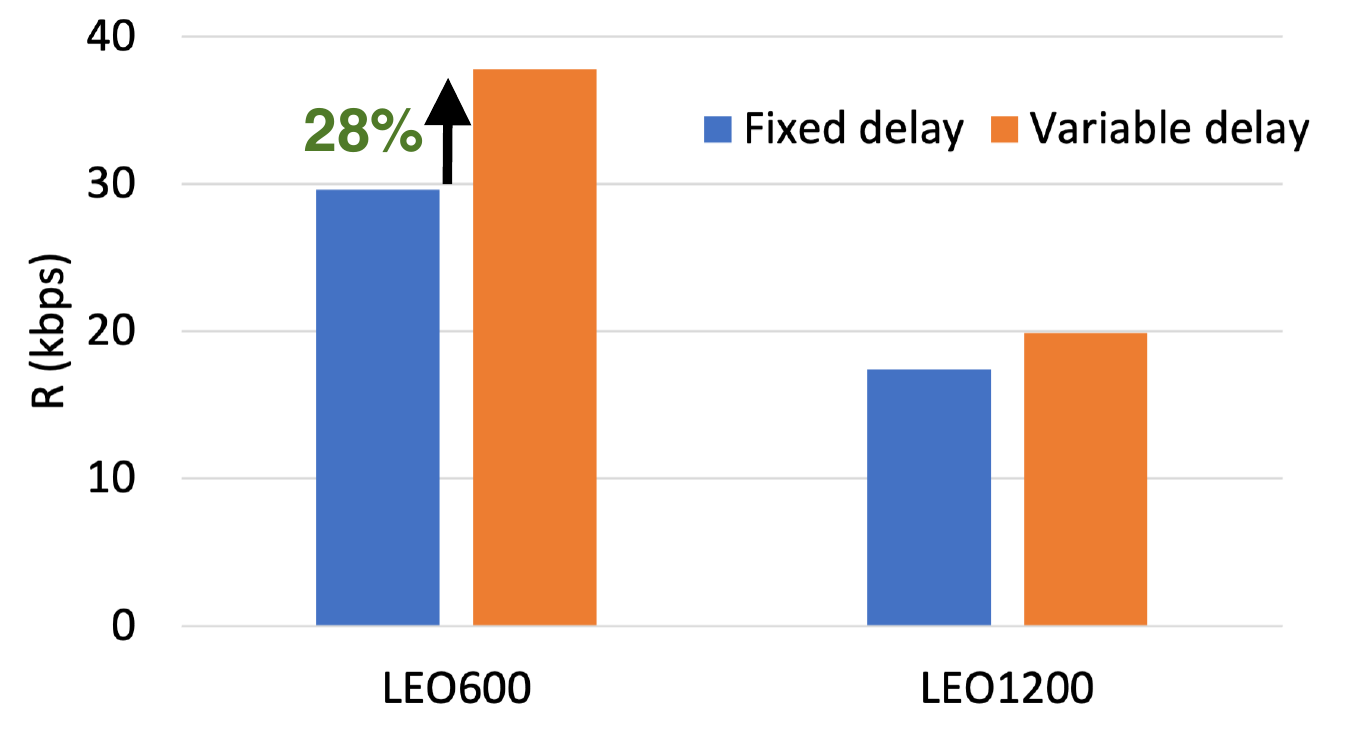} 
	\caption{Throughput for LEO600 and LEO1200 for the conventional fixed delay method and the proposed variable delay design in an LTE-M UE.}
	\label{fig:TP_MTC}
\end{figure}
\begin{figure}[t]
	\centering
	\includegraphics[width=7cm]{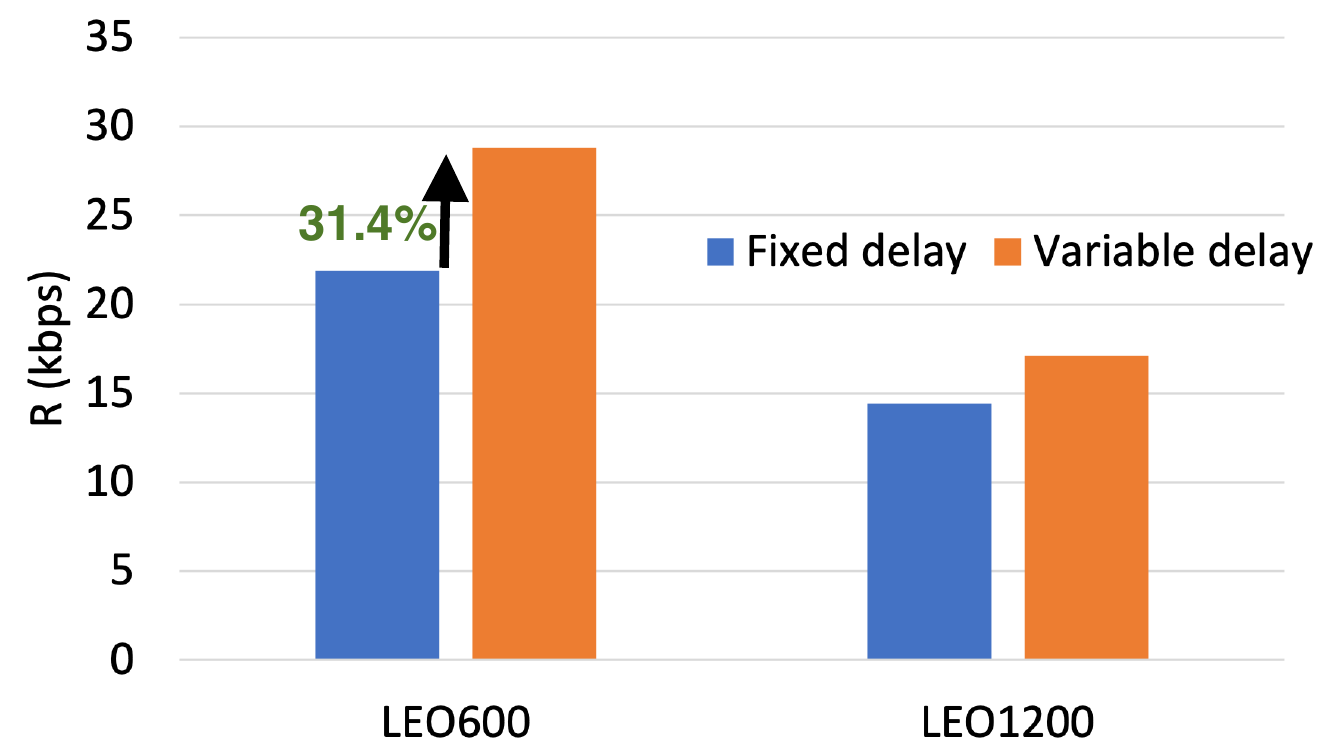} 
	\caption{Throughput for LEO600 and LEO1200 for the conventional fixed delay method and the proposed variable delay design in an NB-IoT UE.}
	\label{fig:TP_NBIOT}
\end{figure}

Next, we compute the throughput using~\eqref{eq:suf_ul} and~\eqref{eq:throughput_general}. We begin with LTE-M systems, where the state-of-the-art method uses a fixed $n_\text{UG2D} = 3$ and one switching SF. \textcolor{black}{We compare the achievable data rates for the state-of-the-art technique using a fixed delay scheduling method and our proposed solution with variable $n_\text{UG2D}$ in Fig.~\ref{fig:TP_MTC}.}   The results demonstrate that we achieve a $28\%$ increase in data rate using our proposed method for LEO600 satellite constellation altitude. For the case of NB-IoT UEs, whose results are shown in Fig.~\ref{fig:TP_NBIOT}, we observe an even higher increase in throughput of over $31\%$ for LEO600 with the use of $n_\text{UG2D} = 8$ for the fixed delay design and two SFs allotted for switching the UE between transmission and reception. Note that the results for NB-IoT were with an increased $N_\harq = 4$, whereas the LTE-M system evaluations were with the current allowed maximum of $N_\harq = 8$.

We further observe in both Fig.~\ref{fig:TP_MTC} and Fig.~\ref{fig:TP_NBIOT} that the throughput gains obtained with the use of our method is higher for lower satellite altitudes. This is because the value of $n_\rep$ required to achieve a target BLER increases with higher satellite altitudes due to the increase in pathloss that demands lower code rates. As a result, the increase in throughput obtained with greater number of TBs per HARQ cycle crosses a point of diminishing returns. This result is also intuitive, since a higher number of repetitions can cover a larger portion of the propagation time delay and therefore presents smaller stop-and-wait gaps to be filled with the use of a larger number of TBs per HARQ cycle. Consequently, this reduces the amount of throughput gain that is achievable from our proposed solution. Nevertheless, for each of these cases, our method provides a superior value of $R$ by reducing the transmission overheads associated with HARQ-based communications. 

\subsubsection{\textcolor{black}{Run-time Complexity and Power Consumption}}
\textcolor{black}{In our final evaluation portion, we present computational complexity results associated with the use of our solution. In particular, we show the run-time complexity and the power consumption resulting from our solution. To this end, we only focus on the computations performed at the battery-powered UE side. }

\begin{figure}[t]
	\centering
	\includegraphics[width=8.5cm]{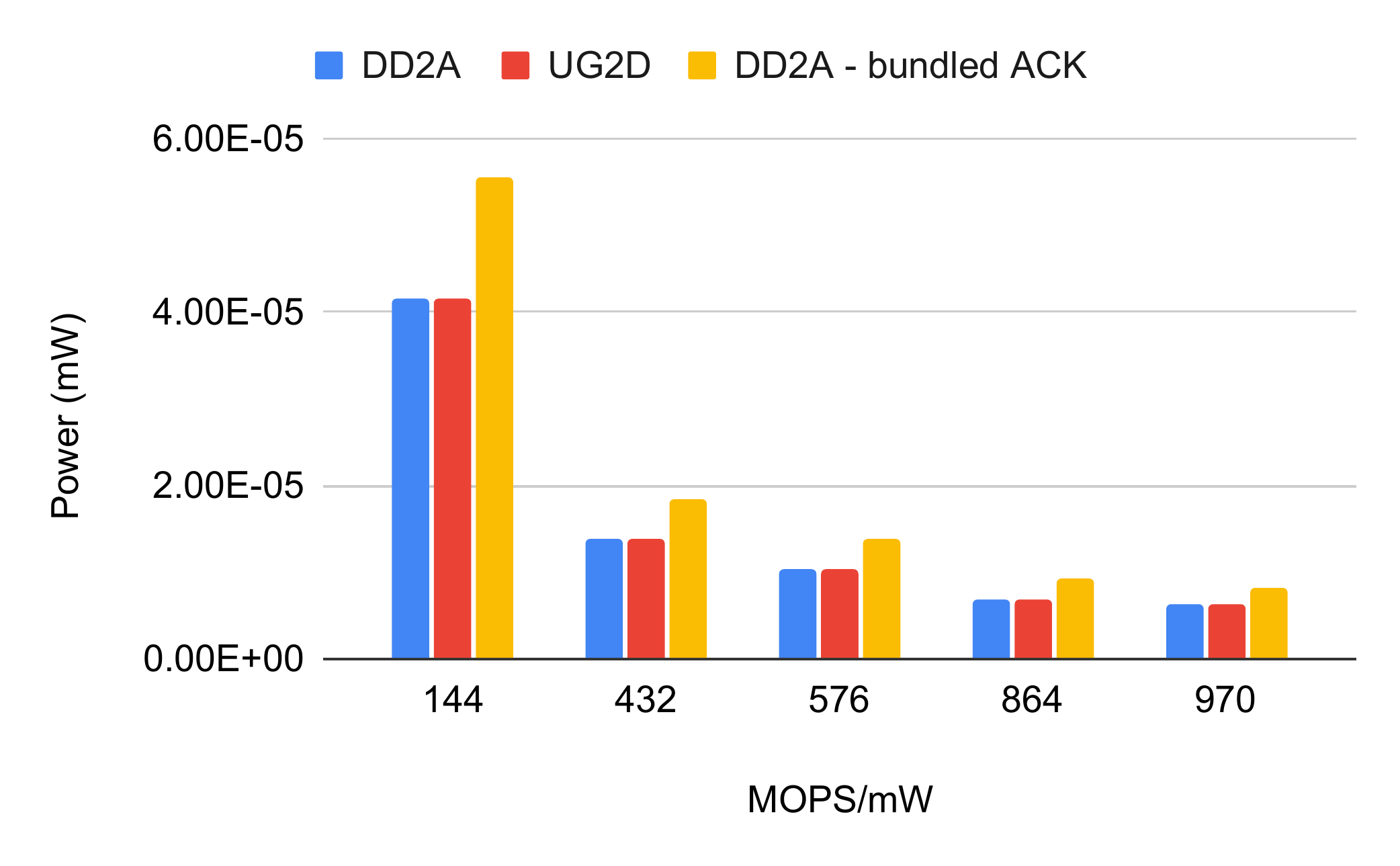} 
	\caption{\textcolor{black}{Power consumption for implementing our solution in a UE with different processor architectures.}}
	\label{fig:chart_comput}
\end{figure}

\textcolor{black}{The additional computations at the UE-end associated with the deployment of our solution include calculating the variable delays shown in Section~\ref{sec:proposed_solutions}. We first determine the number of operations involved with computing DD2A and UG2D. We then use a worst-case assumption that the computations are performed at every SF level, i.e., once every $1$~ms. Note that this is an exaggerated case. In practice, computations at the UE-side need to be performed only when there is a noticeable change in the RTT, for example, when the BS uses a different $N_{\text{TBPHC}}$ in response to the RTT variation. Previous investigations have shown that the RTT varies at a rate of less than $100~\mu$s/s for a LEO satellite at an altitude of  $600$~km and a UE that is moving at a speed of $1200$~km/h in the opposite direction of the satellite movement~\cite{nokia_NTN}. This results in an RTT variation of less than a nanosecond per SF.}

\textcolor{black}{We compute the additional power consumptions for each of our adaptive delay computation methods (i.e., DD2A, UG2D, and DD2A with bundled ACK) as}
\begin{equation}
    P_\phi = R_\text{ops}\frac{\sigma_\phi}{\eta_\text{proc}},
\end{equation}
where $P_\phi$ is the power consumption associated with computing the variable delays for a scheme represented by $\phi \in \{\text{DD2A, UG2D, DD2A-bundled ACK}\}$, $\sigma_\phi$ is the number of operations (i.e., additions, subtractions, multiplications, divisions, and floors) in millions determined from \eqref{eq:dd2a_mtbg}, \eqref{eq:ug2d_mtbg}, and \eqref{eq:ackbundle} for $\phi=\text{DD2A}$, $\phi=\text{UG2D}$, and $\phi=\text{DD2A-bundled ACK}$, respectively, $R_\text{ops}$ is the rate of computations, which we set as once every $1$~ms as described previously, and $\eta_\text{proc}$ is the processor efficiency in million operations per second (MOPS) per mW.

\textcolor{black}{We present the results of $P_\phi$ in Fig.~\ref{fig:chart_comput} for different values of $\eta_\text{proc}$~\cite[Ch. 5]{fasthuber2013energy}. The results demonstrate that our method introduces less than $60~$nW of additional power, even with a relatively less efficient processor that provides an efficiency of $144$~MOPS per mW. With a superior processor efficiency, e.g., $970$~MOPS/mW, power consumption of our proposed soltuion drops to less than $7$~nW to determine the variable delays.}

\textcolor{black}{We put these numbers in perspective by comparing them against typical operating power consumed by commercial low-power IoT UEs. To this end, we use the measurement results obtained for power consumption in NB-IoT devices~\cite{lauridsen2018empirical}. NB-IoT UEs are shown to consume at least $200$~mW and $700$~mW of power in the uplink and downlink, respectively. The maximum of the power consumption numbers from Fig.~\ref{fig:chart_comput} shows that adaptive DD2A with bundled ACK and UG2D computations using our proposed method introduces an additional $60$~nW and $40$~nW, respectively. It can be seen that these numbers are several orders of magnitude lower than the overall power consumption of the device. Such a low power consumption introduces a truly negligible impact on the battery life of an IoT device. }

\section{Conclusion}\label{sec:conclusion}
We proposes a dynamic HARQ scheduling design targeted at IoT-NTN UEs to exploit the extended signal propagation time encountered in satellite communication links. We presented an analysis to determine a suitable number of HARQs to be supported to extract superior throughput under any propagation condition. Our detailed simulation evaluation demonstrated noticeable gains in the achieved throughput with the use of our proposed methods, by considering suitable coding rates for different transmission link conditions and satellite altitudes. Our solution enables NTNs to serve an increased number of UEs, which is critical, given the extended cell-size in NTNs and the increasing number of interconnected devices.

\bibliographystyle{IEEEtran}		
\bibliography{References}{}


\end{document}